\documentclass[twocolumn,aps,prb,showpacs]{revtex4}

\usepackage{graphicx}
\usepackage{dcolumn}
\usepackage{bm}
\usepackage{color}
\input epsf
\input{epsf.sty}

\begin{document}

\preprint{APS/123-QED}

\title {Magnetization reversal and two level fluctuations 
by spin-injection in a ferromagnetic metallic layer}

\author{J.- E. Wegrowe}
\email{jean-eric.wegrowe@polytechnique.fr}
\affiliation{Laboratoire des Solides Irradi\'es, Ecole 
polytechnique, CNRS-UMR 7642 \& CEA/DSM/DRECAM, 91128 Palaiseau Cedex, France.}

\date{\today} 

\begin{abstract}
Slow magnetic relaxation and two level fluctuations measurements under
high current injection is performed in single-contacted ferromagnetic
nanostructures.  The magnetic configurations of the samples are
described by two metastable states of the uniform
magnetization.  The current-dependent effective energy barrier due to
spin-transfer from the current to the magnetic layer is measured.  The
comparison between the results obtained with Ni nanowires of 6 $\mu $m
length and 60 nm diameter, and Co (10 nm)/Cu (10 nm)/Co(30 nm)
nanometric pillars of about 40 nm in diameter refined the
characterization of this effect.  It is shown that all observed
features cannot be reduced to the action of a current dependent
effective field.  Instead, all measurements can be described in terms
of an effective temperature, which depends on the current amplitude
and direction.  The system is then analogous to an unstable open
system.  The effect of current induced magnetization reversal is
interpreted as the balance of spin injection between both interfaces
of the ferromagnetic layer.

\end{abstract}


\maketitle

\section{INTRODUCTION}

The present paper adresses the probleme of magnetization reversal
provoked by injection of spin-polarized current (current-induced magnetization
switching, or {\bf CIMS}). In the context of the 
emergence of spintronics \cite{0}, the discoveries of 
spin-injection \cite{Johnson,Wyder}, 
giant-magnetoresistance \cite{GMR1,GMR2} and tunneling magnetoresistance 
\cite{TMR} created an 
interest in playing with both the spin degree of freedom of the electrons 
and the usual electronic properties.
The CIMS effect was first
predicted \cite{Berger1,Sloncz}, and observed recently by a series of
measurements
\cite{Sun0,EPL,Myers0,Albert,Grollier,EPL2,APL,JAP,Sun,Myers,Kent,Pratt,Andrea}.  In 
these experimental works, the 
magnetization reversal is ascribed to the effect of the spin-polarized
conduction electrons on the magnetization. Some microscopic mechanisms 
of spin transfer from the spin-polarized current to the magnetic layer
have been proposed 
in the framework of different
formalisms \cite{Bazaliy,Heide,Waintal,Stiles,Bauer,Zhang}.

Both experimental and theoretical approaches focus on the typical
Co/Cu/Co pillar system, nanometric in all dimensions
\cite{Albert,Grollier,APL,Sun,Myers,Kent,Pratt,Andrea}.  This is a
pseudo spin-valve structure where the spacer Cu layer is about 10 nm
and the Co layers vary from 1.5 to 30 nm.  This structure is
convenient because it is composed of a spin-polarizer of the current
("pinned" Co layer of 30 nm), and a magnetic layer which plays the
role of an analyzer ("free" Co layer), so that the spin-polarization
of the current is known.  Furthermore, CIMS was associated with giant
magnetoresistance (GMR), or spin accumulation, and used as a probe
allowing magnetization configurations to be measured.  The magnetic
configuration of the two Co layers is either parallel or antiparallel. 
However, the CIMS effect was also measured on homogeneous Ni
nanowires, with anisotropic magnetoresistance (AMR) as a probe, and
without an explicit spin-polarizer \cite{EPL,EPL2}.  The
characteristics of CIMS in Ni nanowires and Co/Cu/Co pillars are very
similar in all points \cite{JAP}, except, in contrast to the pillar
structure, it was not possible to reverse the magnetization in both
directions at fixed field while inverting the sign of the current in
the Ni nanowires.  This last point could be attributed to the fact
that the corresponding double well potential is too asymmetric to
allow reversal in both directions.  In this picture, the effect of the
current is described in terms of effective temperature, but not as a
current dependent effective field whose characteristic is to bias the
profile of the energy potential.

In order to investigate this hypothesis, I discuss here the results of
the experiments of the response of the magnetization to the current
excitation.  Experiments of slow relaxation are presented in terms of
activation process out of a metastable state of the magnetization due
to spin injection.  The typical time range of the current excitation
is about 0.1 to some tens of microseconds.  At this time scale, the
magnetization reversal (with or without current injection) is an
instantaneous event because the typical time of the magnetization
dynamics (the duration of the irreversible jump of the magnetization)
is below the nanosecond \cite{Carlos,Chappert}.  A the time scale
above some few nanoseconds, the dynamics is defined by the time needed
to overcome the energy barrier due to Brownian motion \cite{Coffey}. 
Slow relaxation measurements (or after-effect measurements) and two
level fluctuations hence allow to access to the energy barrier
separating the two metastable states of the magnetization and to the
profile of the energy potential.  Slow relaxation measurements under
current injection is consequently a direct measurement of the energy
transferred from the current to the magnetic layer.  Furthermore, we
show that it is possible to differentiate between a transfer of energy
due to the action of a (current-dependent) effective field or due to
the action of a stochastic diffusion process or magnetization exchange
with spin-polarized reservoirs \cite{PRBthermo}.  In order to discuss
the effect of precession and spin accumulation (or GMR), we present a
comparative study between two different structures performed with an
identical experimental protocol.  Namely, electrodeposited Ni
nanowires of 6 $\mu $m length and 60 nm diameter, and Co(10 nm)/ Cu(10
nm)/ Co(30 nm) pillar structures electrodeposited in the center of a
Cu wire.  The typical energy transferred of the first sample is about
30 000 K per mA (40 000 K for 10$^{7}$ A/cm$^{2}$)) and about 6 000 K
per mA (2000 K for 10$^{7}$ A/cm$^{2}$) for the second sample.

\section{Samples} 

The samples are obtained by a template synthesis method applied to
polycarbonate nanoporous membranes.  The template
synthesis method is described elsewhere \cite{JAP}.  We start with a nanoporous
membrane, e.g. a polycarbonate membrane, with a random lattice of
parallel pores obtained by ion track technology.  Such membranes are
commercially available, with about 6 $\mu $m thickness and pore
diameters down to about 30 nm.  A metallic layer is deposited on the
top (some few tens of nm in order to avoid blocking the pores) and the
bottom (some few 100 of nm in order to block the pores).  The membrane
is then put into an electrolytic bath.  The electrolytic deposition of
the metals inside the pores forms the nanowires.

It is also possible to control the morphology during the growth by
playing with pulsed electrolytic potentials in a bath with two or
more ions.  A structure of, e.g. one or more magnetic multilayer can then be
deposited in the middle of a Cu wire \cite{APL}.  The potential
between the two membranes is measured during the growth, in order to
stop the deposition (with a feed back loop) as soon as a first contact
is obtained.  A single wire is then contacted to the top electrode
during the growth.  Using this method, I have studied the effect
of current injection in the different structures shown in Fig. 1, from
homogeneous 6 $\mu$m Ni nanowires \cite{EPL} to Co(30 nm)/Cu(10
nm)/Co(10 nm) nanopillars electrodeposited in the center of a Cu wire
\cite{APL}, {\it via} the hybrid structures composed of both a
homogeneous Ni part and a multilayered Co/Cu part \cite{EPL2}.


\begin{figure}[h]

\centerline{\epsfxsize 8cm \epsfbox{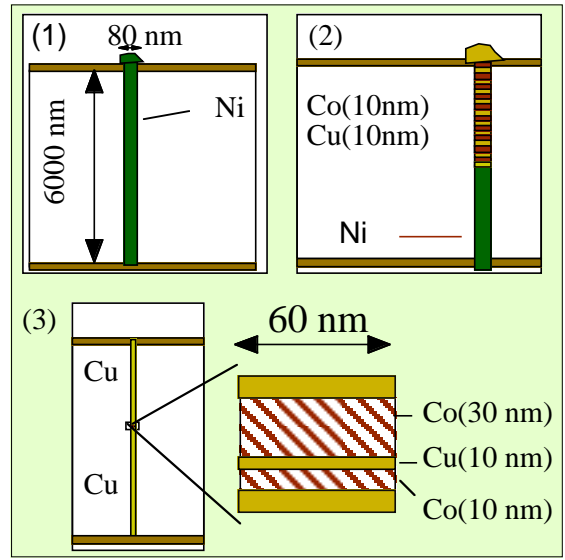}}
\caption{Schematics of three types of single contacted nanowires 
(samples of type A in the text). (1) homogeneous Ni nanowire, (2) 
hybrid structure with a Ni part and a Co/Cu multilayered part and (3) 
Co/Cu/Co nanopillar structure}
\label{}
\end{figure}

In these systems, the structure allows a current to flow
perpendicular to the plane of the ferromagnetic layers (this is the so
called {\bf CPP} geometry).  Because the current is spin-polarized
perpendicularly to the displacement of the electrons in the first
ferromagnetic layer, the CPP geometry enables the spin injection with
well-defined spin polarization in the next ferromagnetic layer. 
However, the spin accumulation, which describes the spin diffusion at
the interface, smooth out the spin polarization in both sides of the
interface \cite{Johnson,Wyder,Fert}.  The spin diffusion length of
electrodeposited Cu and Co is of the order of some few tens of
nanometers.  As will be described below, this spin-polarization allows
the magnetization state to be observed with GMR measurements. 
However, beyond the GMR effect the spin injection may also lead to
CIMS effect as evidenced by the magnetization reversal.  In the case
of homogeneous Ni nanowires, there is no explicit spin-polarizer, and
no GMR can be measured.  Instead, the magnetic configuration is
measured by anisotropic magnetoresistance ({\bf AMR}).  However, also
in this case CIMS effect can be observed, as shown below.  The typical
features discussed in this report have been reproduced an many samples
of each kind, by varying the diameter and the length of the layers,
and the resistance of the contacts.

\section{Spin-injection induced magnetization reversal}


I shall try here to describe the
observed effects from a phenomenological point of view of the 
measured
magnetization reversal process, without introducing any
hypothesis about the microscopic mechanism involved in the transport processes.

The first interest in working with magnetic nanostructures is to be
able to measure a single magnetic domain.  The typical size must hence
be below the typical domain wall size, which is around 100 nm for Ni
and 10 nm for Co.  One can check experimentally that the samples are
indeed single magnetic domain \cite{PRL} (if the magnetocrystalline
anisotropy is weak, the stray field, or shape anisotropy field
forbids the creation of domains perpendicular to the wire; this is
the case in Ni samples (1) and (2) of Fig. 1).  We manage furthermore, by
selecting our samples, to have a uniaxial anisotropy in the Co layers.  Under these
assumptions the magnetic energy can be written in the following form :

\begin{equation}
E = K \sin^{2}(\varphi)+M_{S}H \cos(\varphi - 
\theta)
\label{potential}
\end{equation} 

where K is the anisotropy constant (which includes shape anisotropy),
$ \varphi $ and $ \theta $ are respectively the angle of the
magnetization direction and the angle of the magnetic field $ H $ with
respect to the anisotropy axis, and $M_{s}$ is the magnetization at
saturation.  Note that $ H $ contains all components of the effective
field (see below), except the anisotropy due to its quadratic
dependence.  This function displays a double-well potential as a
function of the magnetic coordinate $ \varphi $ (Fig. 2).

   \begin{figure}[h]
\centerline{\epsfxsize 8cm \epsfbox{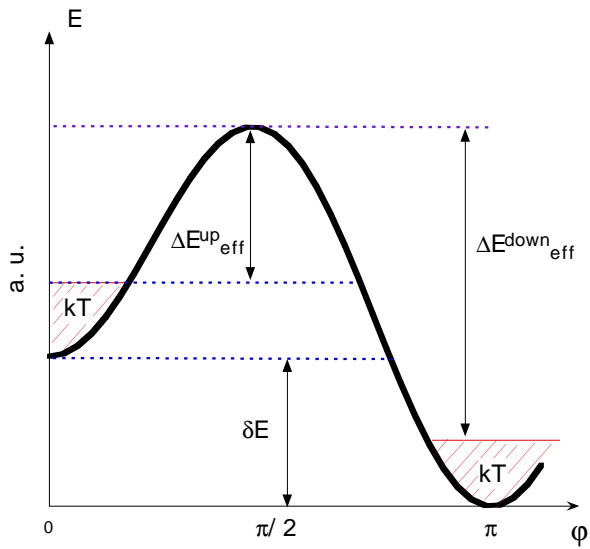}}
\caption{Double well potential (continuous line) with 
stochastic fluctuations (dashed area), and the definition of the 
effective barrier heights and asymmetry.}
\end{figure}

The double well structure of the energy is the archetype of the
hysteretic behavior.  The critical field $ H_{sw} $ defined by a
vanishing barrier height, describes an irreversible jump (and
corresponds to the so-called spinodal limit).  This corresponds to an
irreversible switch of the magnetization from the position defined by
the first energy minimum, to the other equilibrium position around $
\varphi = \pi $.  This irreversible process occurs at $H \approx
H_{sw}¥$ in a time scale below one nanosecond \cite{Carlos,Chappert}. 
A change of the parameter $ H$ (at fixed $\theta $) with $ H \leq
H_{sw} $ corresponds to a reversible rotation of the magnetization.  A
change in the parameter $ H $ with $ H_{sw} \leq H$ does not change
the equilibrium position.  The hysteresis is then composed by a
reversible part and an irreversible part reduced to the magnetization
reversal.  The hysteresis is totally symmetric with respect the
coordinate axes.  The magnetic hysteresis loop for the Ni nanowire is
measured through the anisotropic magnetoresistance (AMR) effect (Fig. 
3(a)), at different angles $\theta$ \cite{PRL}.  The typical amplitude
of the AMR in Ni is about 2 \% of the resistance.  In contrast, the
hysteresis loop for the pillar structure is measured by the GMR effect
(Fig. 3(b)) of amplitude 20 to 40 \% of the resistance of the active
part of the sample (the Co/Cu/Co layers represent about 1 \% of the
total resistance in sample (3) of Fig. 1).  The reversible rotation
of the magnetization is easy to see in the AMR response at large angle
of the applied field ($\theta$ about $80^{\circ}$) in Fig 3(c).  The
GMR of the pillar is measured with the applied field close to the
anisotropy axis in the plane of the Co layers ($\theta$ about
$10^{\circ}$).  The diameter of the Cu wire is now about 40 nm (37
$\pm$ 3), in order to maintain single domain behavior.  The
anisotropy is inside the plane of the layer, as shown by the minor
loop plotted in Fig 3(d) \cite{JAP}.  The current injection is
performed at fixed external field, at a given distance $ \Delta H$ to
the irreversible switch $H_{sw}$, and is represented in Fig.  3(c) and
Fig.  3(d) by the arrows.  The maximum distance $ \Delta H$ from the
switching field with a current density of $10^{7} A/cm^{2}$ is about
50 mT for the Ni nanowire (40\% of the switching field) and 32 mT for
the pillar structure (80 \% of the switching field).

\begin{figure}[h]
\centerline{\epsfxsize 8cm \epsfbox{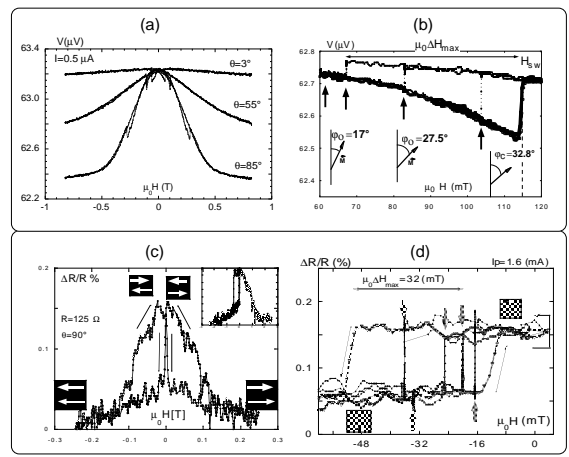}}
\caption{AMR hysteresis loop of sample of type (1) measures at different angles 
of the applied field.  (b) Detail on the irreversible part of the 
hysteresis loop with the effect of current injections (arrows).  The 
magnetization states are sketched.  (c) CPP-GMR hysteresis loop of a 
pseudo spin-valve, with 
minor loop in the inset.  (d) Minor loop with current injection 
(arrows) for the two transitions. }
\label{}
\end{figure}

The equation of the movement of the magnetization is described by the 
Landau-Lifshitz-Gilbert (LLG) equation : 

\begin{eqnarray}
\frac{d\vec{M}}{dt}\,&= &\,g'M_{s}\left(\vec{M} \times 
\vec{H}_{eff}
\right) \nonumber \\
&+ & h' \left(\vec{M}\times \vec{H}_{eff} \right) \times 
\vec{M}
\label{LLG}
\end{eqnarray} 

where g' and h' are constant related to the gyromagnetic ratio and to
the Gilbert damping coefficient. This equation says that the variation
of the magnetization is always perpendicular to the magnetization,
whatever the effective field $ \vec{H}_{eff}$, and is composed by a
precessional term (including transverse relaxation), and a
longitudinal relaxation (second term in the right hand side of the
equation).  

The equation (\ref{LLG}) is deterministic.  The effective field plays
the role of a generalized force, thermodynamically conjugated to the
Gibbs magnetic energy $ E $ \cite{PRBthermo},
 
 \begin{equation}
 \vec{H}_{eff} = - \vec{\nabla}_{M} E
 \label{Heff}
 \end{equation} 

so that $ \vec{H}_{eff}$ contains the anisotropy field $ \vec{H}_{a}$ , the
applied field $ \vec{H}_{ap}$ , the dipole field $ \vec{H}_{d}$ produced by the
presence of the other layer (sample (3) of Fig. 1) if necessary :
$ \vec{H}_{eff} \, =\,\vec{H_{a}}+\vec{H} \, =\,
\vec{H}_{a}+\vec{H}_{ap}+\vec{H}_{dip}$. 

However, as far as we are working
with nanostructures, the fluctuations are of fundamental importance,
and a stochastic term must be added to the LLG equation.  We obtain
the (non-linear) rotational Langevin equation, or equivalently the
rotational Fokker-Planck equation of the probability density $ W $ which
describes a rotational Brownian motion \cite{Coffey}.

\begin{eqnarray}
\frac{\partial W}{\partial t}\,&=& \,g' \, \vec{u} \cdot \left( \vec{\nabla}_{M} W \times 
\vec{H}_{eff} \right) \nonumber \\ 
& + & h' \, 
\vec{\nabla}_{M} \cdot \left ( W  \vec{H}_{eff} \right ) + k' \, 
\nabla^2 _{M} W
\label{FPE}
\end{eqnarray} 

where the first term in the right hand size is the precessional term,
the second term is the longitudinal relaxation, and the third term is
the diffusion term.  The constant k' is evaluated by requiring that
the Maxwell-Boltzmann distribution of orientations is the equilibrium
solution of the energy minima.  In order to analyze after-effect experiments (i.e. slow
relaxation measurements), the activation process over the potential
barrier is described with the relaxation time (or
the first passage time) corresponding to an exponential
relaxation.  Since the precession occurs typically at some tens of picoseconds
(1 to 100 picoseconds), as measured by ferromagnetic
resonance, {\it the precessional term is neglected } \cite{Coffey} (
this approximation will be discussed latter).  The typical 
N\'eel-Brown law
is then obtained:

\begin{equation}
\tau =\tau_{0} \exp \left (\frac{E_{0}(1-H/H^0_{sw})^{\alpha}}{kT} 
\right )
\label{activation}
\end{equation}

where $ H^0_{sw}$ is of the order of the switching field at zero 
temperature, $\tau_{0}$ is the waiting time (10 ps to 1 ns) related to 
k', $\alpha$ is very
close to 1.5 for all angles $\theta$ , except at zero and $\pi$ angles
where it is equal to 2.  The validity of this relaxation law has
been tested on many magnetic nanostructures, including our 
electrodeposited Ni
nanowires \cite{WW}.

 However, if the asymmetry of the
double well potential is small with respect to the barrier height, the
probability of jumping back to the initial state is important.  The
activation process is now described by the two relaxation times back
and forth \cite{Coffey}

\begin{eqnarray}
\left\{\begin{array}{lll}
\tau^{up} &=& \tau_{0up} e^{\left (
\frac{\Delta E^{up}}{kT}\right )} \\

\tau^{down} &=& \tau_{0down}e^{\left (
\frac{\Delta E^{down}}{kT}\right )}
\label{activation2}
\end{array}\right.
\end{eqnarray}

where, $\tau_{0}$ is the waiting time related to the equilibrium
position (local minima) for the parallel (up) and the antiparallel
(down) magnetic configuration of the two Co layers.  The presence of
the two terms in Eq.~(\ref{activation2}) leads to expect a two level
fluctuation (TLF) process during the measurements. The TLF process is 
less appreciated for device application, however it contains more 
information as the single irreversible jump, since both the barrier 
height and the asymmetry of the double well can be deduced.

What happens while injecting currents? The magnetic 
system composed by the magnetic layer must first be enlarged in order 
to take into account the magnetization of the current sources (i.e. some spin polarized 
reservoirs). The energy $ \tilde E$ of this system is: 

\begin{equation}
\tilde{E} = K \sin^{2}(\varphi)+M_{S}H \cos(\varphi - 
\theta) + \epsilon(I)
\label{potential2}
\end{equation} 

where $\epsilon(I)$ is the energy of the spin-dependent current
source.  A generalized effective field $ \vec{\tilde{H}}_{eff}$ is defined
from the energy $ \tilde E$ by the relation:

 \begin{equation}
 \vec{\tilde{H}}_{eff} = - \vec{\nabla}_{M} \tilde {E}
 \label{GHeff}
 \end{equation} 

and this effective field contains an other term which includes the
effect of the current : $ \vec{\tilde{H}}_{eff}=
\vec{H}_{a}+\vec{H}_{ap}+\vec{H}_{dip}+\vec{H}(I)$.  The {\it exchange torque} term
\cite{Sloncz} of the
form $\vec{M}\times \vec{H}(I)$ where $
\vec{H}(I)=Ia(\vec{s}.\vec{M})(\vec{s} \times \vec{M}) $ and $ \vec{s}
$ is the spin polarization of the current must necessarily be
completed by an other term to give $ \vec{\tilde{H}}_{eff}$, because it cannot be derived from a
potential function.  My goal is to present a
phenomenological approach (based on the measured macroscopic 
magnetization $ \vec{M}) $) and
the link to microscopic approaches \cite{Berger1,Sloncz,Bazaliy,Heide,Waintal,Stiles,Bauer,Zhang} is
beyond the scope of the present work \cite{Rque}. Hence we write the 
generalized LLG equation,  without any loss of
generality, in the following
form :

\begin{eqnarray}
\frac{d\vec{\tilde{M}}}{dt}\,&=& \,g'M_{s}\left(\vec{M} \times 
\vec{ \tilde{H}}_{eff}
\right) \, \nonumber  \\ 
&+& \, h' \left( \vec{M} \times \vec{\tilde{H}}_{eff} \right ) \times 
\vec{M}\,+\, f(I,\vec{M}) \vec{M} 
\label{GLLG}
\end{eqnarray} 

The third term in the right hand side of 
equation (\ref{GLLG}) is
due to the so called {\it longitudinal spin transfer} which does not conserve the
magnetization of the ferromagnetic layer. $f$ is a 
function of the current and the magnetization configuration. This terms 
includes all possible mechanisms which lead to non conservation of the 
magnetization, including generation of spin-waves or magnons due to 
the current \cite{Tsoi}, or any relaxation channel from the spins of 
the conduction 
electrons to a ferromagnetic collective variable $\vec M$. The other components of the
spin-transfer (or Oersted fields) are included into the effective
field.

In order to summarise, the usual N\'eel-Brown activation process (slow relaxation) 
allows one to access to the effective potential profile, and hence the 
current-dependent effective 
field. Since the third 
term added to the LLG equation and describes in (\ref{GLLG}) does 
not act on the potential profile, it is contained in the 
non-deterministic part of equation 
Eq.~(\ref{FPE}). The 
stochastic fluctuations hence include the thermal fluctuations kT, 
the magnon generation, and any other relaxation channels from the 
spin-polarized current to 
magnetization. The corresponding energy (dashed parts in Fig. 2) 
 defines the measured effective barrier heights $\Delta E_{eff}^{up}$ and $ \Delta 
 E_{eff}^{down}$.  In the absence of a more detailed stochastic 
theory of activation due to spin transfer, it can conveniently  be 
accounted for by a phenomenological effective temperature $T_{eff}(I)$ 
such that $E_{eff} = k_{B}T_{eff}(I)$. This temperature is expected to play a different 
role depending on the current sense, and is also expected to depend 
on the magnetic configuration. 

In order to analyze the data presented below, let me first suppose
that the third term in Eq.~(\ref{GLLG}) is not present.  This
means that the system composed by the ferromagnetic layer is supposed
to be closed (though non adiabatic due to the current injection),
because the third term describes an open system.  Then, the derivation
of the activation process is unchanged, and we expect to measure the
same relaxation process as described by equation (\ref{activation}) with
$H$ replaced by $\tilde{H}$ (and $\theta $ by $\tilde \theta $).

\section{Experimental results}

In this study, the response to the injection of current is measured at
the time scale 0.1 to 10 microseconds.  The time scale is chosen in
order to avoid any non-stationary heating regimes.  The measurements
are performed using a Wheatstone bridge, and a Lecroy Gigasampler (the
experimental set-up and the thermal regimes are discussed elsewhere
\cite{Carlos}).  The external field is first set at saturation ($\pm$
1 Tesla), and then decreased to a fixed value $H$.  The current is
injected with a step function at time t=0.  The raising time of the
current injection was changed between 0.01 to 1 microsecond (i.e.
varying the cut-off frequency of the current excitation between 1 and
100 MHz) without significant change in the response of the
magnetization.  A statistical assessment is performed over many
measurements, by cycling the hysteresis loop with the external field
before each measurement.  This protocol is repeated for different
values of the applied field $H$, and different values of the amplitude
of the current excitation $I$.  The results are presented in Fig.  4
for the Ni sample (sample (1) of Fig 1) and Fig.  5 for the pillar
sample (sample (3) of Fig 1).

{\bf Ni nanowires}.  Figure 4(a) shows the response of the resistance
to the current excitation for a Ni nanowire.  The first change in the
voltage is due to the Joule effect, and a quasi-stationary regime is
reached after about 1 microsecond, at a temperature of 330K. The small
jump in the middle of the curve is the anisotropic magnetoresistance
(AMR) response of the magnetization to the current injection.  The
jump is magnified in Fig.  4(b).  Statistics of the switching time
over many events allows one to get the exponential relaxation by
integrating the histogram over the time.  The mean relaxation time is
then deduced, and plotted as a function of the external field for
different currents in fig.  4(c).  The fit of the curves are performed
with the N\'eel-Brown activation laws Eq.~(\ref{activation}), with the
current dependence contained entirely in the energy term $E_{0}$.  The
curves of Fig.  4(c) cannot be accounted for by a current dependent
field $H(I)$ \cite{JAP}.  Instead, there is an energy variation as
plotted in Fig.  4(d).  If linearised (see e.g. the other sample
reported in \cite{Carlos}) $\Delta E_{0} = a' \cdot I$, with $a' $=
30000 K/mA (about 40 000 K/(10$^{7}$ A/cm$^{2}$) ).  Normalized to the
anisotropy energy, the variation of the energy is about 30 \% / mA.
The energy of 30 000 K ($2 .  10^{-19}$ J) corresponds also to $
\Delta H_{max} = 50 mT $ illustrated in Fig.  3(b)
\cite{EPL2,WW}.

\begin{figure}[h]
\centerline{\epsfxsize 8cm \epsfbox{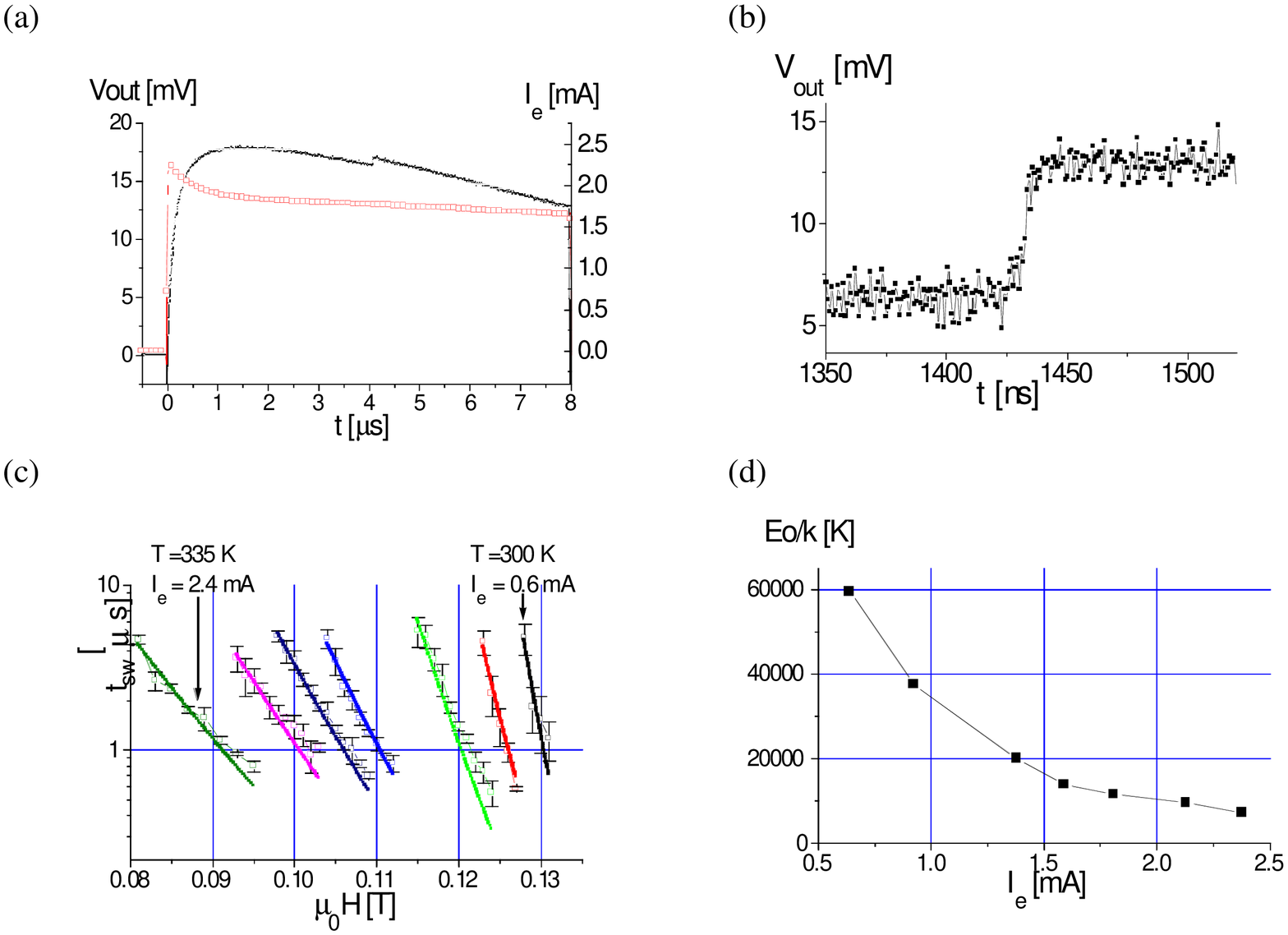}}
\caption{After-effect measurements with homogenous Ni nanowires.  (a) current 
injection (right scale), and response of the resistance (left scale).  
(b) Detail of the AMR response.  (c) Mean switching time as a function 
of the applied field for various current injections.  (d) Variation of 
the parameter $E_{0}$ as a function of the current amplitude.}.
\label{NiAF}
\end{figure}

{\bf Pillar structure}.  In order to insure single Co magnetic domain
behavior, the pillar structure is electrodeposited in a wire of 37 nm
$\pm$ 3 nm in diameter, instead of 70 to 80 nm for the Ni nanowires. 
As a consequence, a current of 0.4 mA corresponds to about $10^{7}$
A/cm$^{2}$.  The same current density required an injection of about
1.5 mA in the Ni wire.  The same protocol was applied to pillar
structure (sample (3) of Fig.  1).  The volume of the Co layer is
about 400 times smaller than that of the Ni wire.  The anisotropy
energy is between 50 to 100 times smaller, depending on the
crystallinity.  The asymmetry of the double well is small with respect
to the barrier height so that the probability of jumping back to the
initial state is important.  The activation process is now described
by the two relaxation times back and forth described in
Eq.~(\ref{activation2}).  The TLF feature is indeed measured {\it
under current injection only}, as shown in Fig.  5, for a large range
of field and current.  The TLF is observed over 0.1 T for the explored
positive and negative currents.  The amplitude of the jumps
corresponds to the GMR signal (one Ohm) between the parallel and
antiparallel magnetic configurations \cite{Andrea}.  The TLF behaviour
was measured in other simmilar systems \cite{Myers,Pratt},
and seems to be the main signature of the CIMS pseudo spin-valves.

\begin{figure}[h]
\centerline{\epsfxsize 8cm \epsfbox{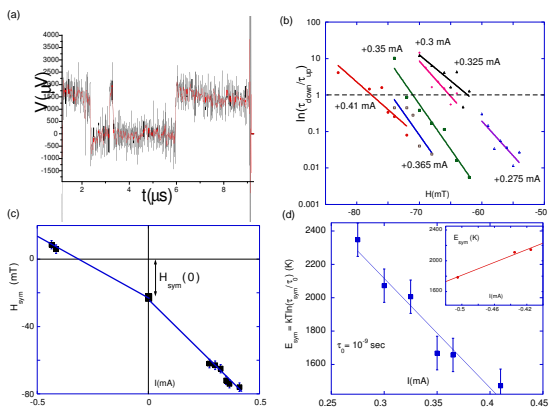}}
\caption{ After-effect measurements measured on the
Cu/Co/Cu pillar.  (a) response of the voltage to the current 
injection of O.4 mA (about $10^{7}$ A/cm$^{2}$).  (b)
Asymmetry of the double well $\delta_{\varphi}E$ as a function of the applied field for 
various current injections.  
(c) Values of the field as a function of the current for symmetric 
double well (defined by $\delta_{\varphi}E=0$). (d) Variation of the 
barrier height $\Delta E (H_{sym})$ of the symmetric double well as a function 
of the current.}
\label{}
\end{figure}

The TLF allows one to access two parameters, the mean time spent in
the antiparallel configuration $\tau_{up}$ and the mean time spent in
the parallel configuration $\tau_{down}$.  The ratio, plotted in Fig. 
5(b) for positive currents gives the parameter $ \delta E(H) $, i.e.
the variation of asymmetry of the energy profile.  The value
$ln(\tau_{down}/\tau_{up})=1$ corresponds to the symmetric double
well.  Note that different values $\tau_{0up} \ne \tau_{0down}$ should
be expected but the difference is negligible with respect to the
exponential behavior.  More important is the possibility of different
{\it effective temperatures} between both sides of the barrier,
especially as a function of the current direction.  This possibility,
and its physical meaning, will be discussed in the next section.  Under
this last assumption, it can be seen that all happens as varying the
current at fixed field were equivalent to bias the energy profile. 
The effect of the current would here be equivalent to the action of a
field.  Tunning the current can be exactly compensated by tunning the
external field, in order to keep the energy profile unchanged (i.e.
following an horizontal line in Fig.  5(b)).  Except for pathological
cases (the spinodal limit), keeping constant the energy profile means
keeping constant the effective field $ \vec{\tilde{H}}_{eff}¥$:

\begin{equation}
E(\vec{\tilde{H}}_{eff})= cst \mapsto \vec{H}_{ap} 
+\vec{H}_{a}+\vec{H}_{d}+\vec{H}(I)=\vec{cst}
\label{E(I)}
\end{equation}

This equation defines the function of the current $H_{ap}=H_{sym}(I)$,
plotted in Fig.  5(b) for a constant profile
($ln(\tau_{down}/\tau_{up})=1$) corresponding to the symmetric double
well.  The symmetric double well at zero current is located in the
middle of the minor hysteresis loop (see \cite{APL}) for the external
field $H_{sym}(0)=-22$ mT. The current dependent effective field is
linear in current, H(I)=$a$I, with a coefficient of the order of
$a$=0.1 T/mA  or 33 mT/(10$^{7}$ A/cm$^{3}$).  The coefficient $a$ is of
the same order of magnitude of what has been measured in previous
studies in terms of critical currents $I_{c}(H)$, where the Co layer
(the analyzer) was 3 to 5 times smaller \cite{Albert,Grollier,Sun}. 
This analysis is based only on th existence of a current dependent effective field
\cite{Sloncz,Berger1}.  However, the TLF effect also gives access to
the value of the energy barrier $ kT \cdot
ln(\tau_{up}(H_{sym}))=\Delta E_{sym}(I) $ as a function of the
current for a fix energy profile (more precisely at a fixed ratio
$\tau^{down}/\tau^{up}$ in Fig 5(c)).  The result is plotted in Fig
5(c), for the symmetric double well profile.  The variation as a
function of the current amplitude is also very important and
corresponds to 1000 K for a variation of 0.15 mA. The variation of the
barrier height is approximately linear $\Delta E_{sym}(I) = a' I $
with a'=6800 K/mA (more than 0.5 eV/mA) for positive current, and
about 4000 K/mA for negative current (plotted in the inset of Fig. 
5(c) and reported in \cite{Andrea}).

\begin{figure}[h]
\centerline{\epsfxsize 8cm \epsfbox{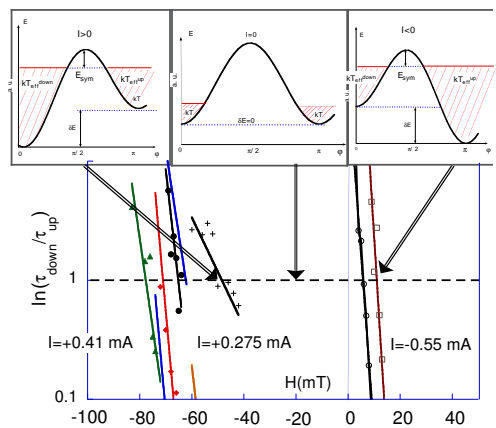}}
\caption{Illustration of the two level fluctuation under current 
injection with the effect of the 
current described exclusively by the phenomenological effective temperatures $ 
T_{eff}^{up}$ and $ T_{eff}^{down}$ (see text). Potential profile for 
symmetric relaxation. From left to right: 
(a) profile of the potential for current injection +I at H=-50 
mT, (b) profile at H=-20 mT without current injection, (c) 
profile at H=+15 mT with current injection -I.}
\label{Teff}
\end{figure}

The current dependence at fixed energy profile cannot be explained
only with the current dependent effective field $\vec{\tilde{H}}_{eff}$
defined in Eq.~(\ref{Heff}).  Another mechanism is necessarily
present.  One possibility is the effect of the precession, i.e. a
resonance, maintained in a stationary regime due to the current
injection.  The second possibility, which includes phenomenologically
the first one (i.e. the precession) is to describe all features
observed with the help of an effective temperature which depends on
the current direction.  In
this picture, the horizontal dashed line $ ln(\tau_{down}/ \tau_{up})=
1$ in Fig.  5(b) represents any double-well energy profile (i.e.
$\delta E \ne 0$ if $I \ne 0$) but with equal effective energy
barriers $ \Delta E_{eff}^{up}= \Delta E_{eff}^{down}$ in the right
and left wells.  The effective temperatures $kT_{eff}^{up} \ne
kT_{eff}^{down} $ are then different in both sides of the barrier. 
This means that the fluctuations (due either to the precession, or to
the longitudinal spin transfer in Eq.~(\ref{GLLG})) are not equal in
the right and left well, and depend on the current direction.  This
interpretation is depicted in Fig 6 together with some of the
experimental data of the TLF measurements.  Three different profiles
of the double well are sketched corresponding to the {\it symmetric
effective barrier height}.  It can be seen that all data are accounted
for within this picture.  At fixed current injection, a line following
the experimental points is due exclusively to the biasing of the
energy profile provoked by the external field.  In contrast, the
variation of the amplitude or sign of the current injection is
described in terms of effective temperature, or in terms of effective
barrier heights.

The hysteresis loop with weak current and high currents of both
directions is shown in Fig.  7.  As already observed in a previous
work \cite{Grollier}, the hysteresis loop is enlarged for positive
current and the hysteresis loop is shrunken for negative current.  On
the basis of the interpretation depicted above, the enlargement of the
hysteresis loop for positive current and the shrinkage of the
hysteresis loop for negative current occurs if the energy
$kT_{eff}^{up}$ or the energy $kT_{eff}^{down} $ is greater than the
maximum barrier height (which corresponds to the symmetric double
well).  It follows that the state of the system is blocked in the
opposite potential well by an effective barrier.  For the antiparallel
alignment, this process can be compensated by the Zeeman energy which
is able to force the two Co layers to align at high enough magnetic
field above the energy $kT_{eff}$ (at about 1 Tesla in our case).  In
the parallel configuration, the dipole filed (which plays the role of
the "exchange biasing" of real spin-valves) cannot be tuned, and there
is no possibility for imposing the antiparallel configuration : the
hysteresis disappears.  This situation can be described as a
directional superparamagnetism, where only one direction of the
magnetization is averaged out, so that the other direction is imposed. 
Note that in this case, namelly if $kT_{eff} $ is larger than the
maximum energy barrier, the effect of the current injection is
equivalent to adding magnetic momenta in a well-defined direction
(instead of increasing the energy of the opposite magnetization
state), and the description in terms of effective temperature (and the
double well picture) is may not be longer relevant.

\begin{figure}[h]
\epsfxsize=8cm 
\centerline{\epsfxsize 8cm \epsfbox{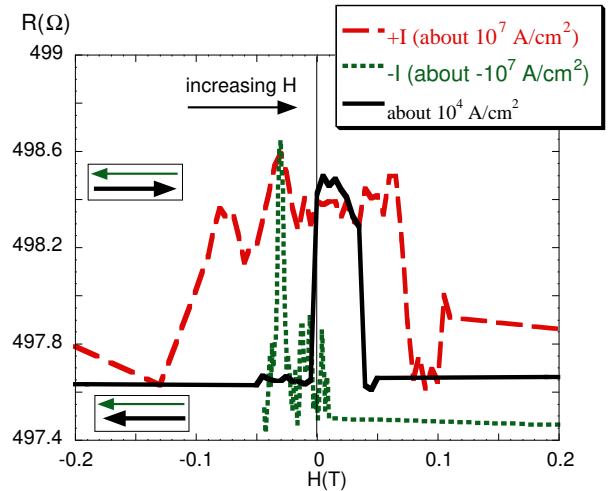}}
\caption{Half hysteresis loop for increasing field
measured with 10 microseconds current pulses (no averaging). Continuous line : GMR 
reference hysteresis loop. Dashed: positive current at about 10$^{7}$ 
A/cm$^{2} $. Dots: negative 
current for the same amplitude.}
\label{HystTLF}
\end{figure}

\subsection{irreversible spin transfer} 

The variation of energy measured under current injection is about 40
000 K/(10$^{7}$ A/cm$^{2}$ (ou 30 000 K/mA ) for Ni nanowires and
about 2000 K/(10$^{7}$ A/cm$^{2})$ (or 6600 K/mA ) for pillar
structures.  More precisely, the results obtained about the activation
process provoked by spin-injection show that a current-dependent
effective field as defined in Eq.~(\ref{GHeff}) and Eq.~(\ref{GLLG}))
is not sufficient to account for the two level fluctuation in 
Co/Cu/Co pillars and to the after-effect measurements in Ni nanowires.  In
contrast, all observed features can be interpreted phenomenologically
without the need to invoke a current dependent effective field, but
with a current dependent effective temperature.

How to interpret this effective temperature?  One possibility is to
store this energy in he form of precession of the magnetization
described by the first term in Eq.~(\ref{GLLG}).  Under this
hypothesis, the energies in the potential wells depicted in Fig 6 are
simply the precession induced by the current (or any coherent
spin-waves).  Such a mechanism has been proposed by Berger
\cite{Berger2} in the term of SWASER (spin-wave amplification by
stimulated emission of radiation) and is being studied by
ferromagnetic resonance \cite{Rezende,Urban}.  Such precession, which
would decrease the amplitude of the resistance jump from one state to
the other has not been observed within the precision of our
measurements (see raw data of Fig 4 (a) and 5 (a)), and this
interpretation would hardly describe the effects observed on the
hysteresis loop.  Furthermore, if it is easy to understand that a high
frequency excitation (beyond the GHz range) is able to maintain a
ferromagnetic resonance (FMR) in a sample of arbitrary size, it is
very hard, in contrast, to justify that a DC current, or a slow step
function (with cut-off frequency below 100 MHz) is able to maintain a
resonance over decades in a {\it macroscopic} dissipative system.  The
comparison between the two systems, Ni and Co /Cu/Co, with aspect
ratio and size difference of more than a factor 100 (from about
10$^{6}$ to 10$^{8}$ coupled spins in the present study), shows that
the effect do not originates from coherent spin-waves. 
However, the consequence of the spin transfer is of course, whatever
its origin, a magnetic excitation which dissipates in the system in
the form of magnetic switching, spin waves, solitons, or uses other
more complicated dispersion channels.

  The other possibility is that the effect is entirely described by
  the third term of Eq.~(\ref{GLLG}) (the so called "longitudinal
  spin-transfer").  The system is then an open system and the
  variation of energy measured under current injection is transferred
  by the injection of magnetic moments at the interface.  In the
  framework of this interpretation, it is necessary to imagine a
  relaxation process from the spin of the conduction electrons to the
  magnetization of the layer.  Both objects are then supposed to be
  described by two separated sub-systems, like $s$ and $d$ electrons
  involving both spin channels \cite{Mott,AMR} (i.e. a four-channel
  model).  This effect would be similar to that of a
  Light-emitting diode, with the difference due to the fact there is 
  probably no well defined
  energy gap, and that the "emission-absorbtion" should be
  partially compensated between both interfaces in case of uniform 
  magnetization.
  
  Without entering in the details of a possible mechanism (see e.g. a
  thermokinetic approach of the four-channel approximation developed
  in reference \cite{fourchannel}) the energy involved may be of the
  order of the band splitting between the two up and down bands, i.e.
  of the order of 0.23 eV for Ni \cite{Himsel,Photolsf}. 
  Consequently, the energy of 30 000 K corresponds to a imbalance of a
  factor 10 in the efficiency of the spin injection between the top
  and the bottom of the wire or the two interfaces of the Co layer. 
  At one interface, the relaxation of electrons from $ s$ to $d$ of
  the minority spin channel leads to add magnetic momenta to the
  magnetic layer, and the effect is inverted at the other interface. 
  If the two interfaces of the magnetic layer are perfectly symmetric,
  the spin injection is compensated at both interfaces.  In contrary,
  if there is an asymmetry of the spin injection at both interfaces, a
  imbalance occurs and a net magnetization is injected in the layer. 
  Consequently, CIMS effect
  should depend crucially on the symmetry of both interface.  This prediction
  is corroborated by the measurements presented in Fig.  8, of a Ni
  nanowire with Ni contacts on one side (and Au on the other side),
  and Ni nanowire with Cu contacts (and Au on the other side).  The
  wire with the Ni contact on one side shows a strong CIMS effect :
  the typical parameter is the slope $a$= 50 mT/(10$^{7}$ A/cm$^{2}$). 
  In contrast, the wires with Cu contacts have a negligible CIMS
  effect, whatever the size of the Ni wire (a set of samples from 5.5
  microns down to 1 micron in length have been measured).

   \begin{figure}[h]
\epsfxsize=8cm 
\centerline{\epsfxsize 8cm \epsfbox{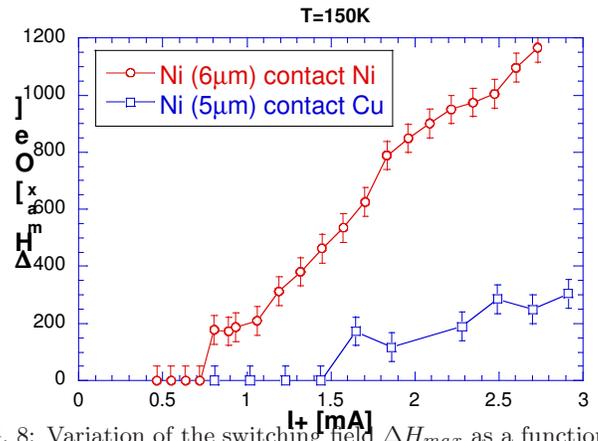}}
\caption{Variation of the switching field $\Delta H_{max} $ as a 
function of the amplitude of the current pulse. The sample with 
asymmetric interface (Ni contact) shows a strong $\Delta H_{max} $ while the 
sample with symmetric interface (Cu 
contact) has a negligible $\Delta H_{max} $.}
\label{Marcel}
\end{figure}

  This interpretation is also corroborated by the counter-intuitive result
  that the amplitude of spin transfer is more important in homogeneous
  Ni than in GMR Co/Cu/Co pillars.  This observation can be understood
  by the fact that spin-accumulation through the 10 nm thick Cu spacer
  layer in the pillar tends to smooth out the spin injection at both
  interfaces and tends to minimize the relaxation.
 
In conclusion, the measurements performed on the different
electrodeposited samples show that the effect of the current cannot be
reduced exclusively to the effect of a current dependent effective
field (more precisely, an effective field thermodynamically defined 
as conjugated to the magnetization). 
If we assume that a macroscopic precession induced by a DC current
during microseconds to seconds is not realistic, one is lead to
conclude that the magnetic layer is an open system with respect to magnetic 
moments.  This means that the
spin-injection transfers magnetic moments from the current to the
magnetic layer.  The microscopic mechanisms responsible for this
transfer, and the relation between the effective field defined here
and the "exchange torque" introduced by Slonczewki and developed in
many microscopic approaches, is still unclear in the absence of a
stochastic theory of the magnetization reversal that includes spin
transfer.  However, the effect can be understood by assuming that the
spin of the conduction electrons and the magnetization are two
distinct quantities, so that some spins are able to relax from one
system to the other, namely from the current to the ferromagnetic
layer, leading to an important transfer in forms of magnetic moments. 
This relaxation can be mediated by magnon excitations, or other types
of magnetic collective variables, but with typical relaxation time
scales measured to be below 25 nanoseconds.


\begin{references}
\bibitem{Rque} In the approach proposed here, the effective field is 
defined
as the intensive variable thermodynamically conjugated to the 
measured magnetization \cite{PRBthermo}. This variable 
may be different to a
current dependent effective field 
defined microscopically.
\end{references}
\end{document}